\documentclass[10pt]{iopart}
\usepackage{placeins}
\usepackage{graphicx}

\begin{document}

\title[Magnetization reversal of a nanocluster]{Thermal and surface anisotropy effects on the magnetization reversal
of a nanocluster}
\author{P.-M. D\'ejardin$^{a}$, H. Kachkachi$^{b}$, and Yu. P. Kalmykov$^{a}$}
\address{$^{a}$ LAMPS, Universit\'e de Perpignan Via Domitia, 52 avenue Paul Alduy, 66860 Perpignan Cedex, France}
\address{$^{b}$ GEMaC, UMR 8635, Universit\'e de Versailles St. Quentin, 45 avenue des Etats-Unis, 78035 Versailles Cedex, France}
\ead{hamid.kachkachi@physique.uvsq.fr}
\begin{abstract}
The relaxation rate and temperature-dependent switching field curve of a spherical magnetic nanocluster are calculated by including the effect of
surface anisotropy via an effective anisotropy model. In particular, it is shown that surface anisotropy may change the thermally activated
magnetization reversal by more than an order of magnitude, and that temperature-dependent switching field curves noticeably deviate from the
Stoner-Wohlfarth astroid. With recent and future $\mu$-SQUID measurements in mind, we indicate how comparison of our results with experimental data on isolated clusters may allow one to obtain valuable information on surface
anisotropy.
\end{abstract}
%
\section{Introduction}

Understanding the dynamics and the mechanisms of the magnetization reversal in magnetic nanoclusters is essential for technological applications especially in the area of data storage. For such applications, rather fine nanoclusters are needed to increase the areal density, within ranges allowed by the superparamagnetic effect. On the other hand, in such small systems the magnetic state at the surface differs in many respects from that in the bulk. As such, the surface/interface effects cannot be ignored in investigations of the dynamics of magnetic nanoclusters. Now, surface effects are local effects and therefore such nanoclusters have to be regarded, in principle, as crystallites of many atomic magnetic moments. The problem with this picture of a nanocluster is that the study of the dynamics becomes a tremendous task because of the large number of degrees of freedom that have to be dealt with when computing, for example, the magnetization reversal time of the nanocluster. A way of overcoming this difficulty has been provided by the development of an effective model \cite{EOSP} which maps the many-spin cluster onto a macrospin representing the net moment of the cluster and whose energy contains mixed uniaxial and cubic anisotropies. The sign and magnitude of the effective coefficients of these anisotropy contributions depend on the size and shape of the cluster, crystal structure of the underlying material, and physical parameters such as the exchange coupling and local anisotropy constants. This model obviously provides a compromise between i) the Stoner-Wohlfarth (SW) macrospin approach [see the review article \cite{wernsdorfer01acp} and references therein] based on coherent rotation of all atomic spins and making use of the N\'{e}el-Brown theory for the calculation of relaxation rates, and ii) the many-spin approach \cite{kacgar05springer,H.B. Braun} employing the Langer's theory of the decay of metastable states \cite{Langer}. In fact, these two theories provide asymptotic relaxation rates for each elementary process, i.e., the transition from a given state to a more stable state through a saddle point, and at this level they are fully identical and yield the same results. Finally, it is worth mentioning that this effective model also makes it possible to take into account more easily inter-particle dipolar interactions and therefore to provide us with a basis for interpreting experimental observations, e.g., through the zero-field-cooled/field-cooled magnetization, AC susceptibility, FMR, etc.

In the present work, we compute the magnetization reversal time and demonstrate the effects of temperature and surface anisotropy (SA) on the switching field of a spherical nanocluster. We also indicate how, by comparing our results with experiments, it may be possible to gain information concerning SA in isolated magnetic nanoclusters.
\section{Energy and relaxation rate}
According to the effective model \cite{EOSP}, the energy of a magnetic nanocluster may be written as
\begin{equation}\label{eq:EnEOSP}
E = -\mu_s\mathbf{H}\cdot\mathbf{m} - K_2V\, m_{z}^{2} + \frac{1}{2}K_4V\sum_{\alpha=x,y,z}m_{\alpha}^{4}
\end{equation}
where $m_{\alpha},\alpha=x,y,z$, with $\vert\textbf{m}\vert = 1$, are the components of the verse of the nanocluster net magnetic moment and $\mu_{s}$ its magnitude, and $V$ the nanocluster volume. $\mathbf{H}$ is the magnetic field applied at some angle $\psi$ with respect to the uniaxial anisotropy (UA) axis $z$. 
As mentioned in the Introduction, the parameters $K_{2}$ and $K_{4}$ are effective anisotropy constants whose sign and magnitude depend on the intrinsic features of the nanocluster: lattice structure, shape and size, the (ferromagnetic) exchange coupling constant  $J$, and on-site core and surface anisotropies. For example, for a spherical cluster with no UA in the core, the cubic anisotropy (CA) contribution in Eq. (\ref{eq:EnEOSP}) is induced by N\'eel's SA and we have $K_{\mathrm{4}}=\kappa\mathcal{N}\, K_{s}^{2}/zJ$, where $\mathcal{N}$ is the number of atoms in the nanocluster, $z$ the coordination number, $K_{s}$ the on-site SA constant, and $\kappa$ ($\simeq0.53465$ here) a surface integral that depends on the intrinsic features mentioned previously and also on the SA model. 
In the sequel, we use the dimensionless parameter $\zeta\equiv K_4/K_2$.

The calculation of the reversal time may be accomplished in two ways. The first is analytical and provides asymptotic expressions
for the relaxation rate corresponding to one escape route, i.e., a passage from an energy minimum $(1)$ to a minimum $(2)$ through a saddle point $(0$). For this one may apply Langer's approach \cite{Langer} which yields the relaxation rate in the form \cite{kac03epl}
\begin{equation}\label{eq:LangerFormula}
\Gamma_{(1)\rightarrow(2)}=\frac{\Omega_{0}}{2\pi}\frac{Z_{0}}{Z_{1}} ,
\end{equation}
where $Z_{0}$ and $Z_{1}$ are the partition functions computed in the neighborhood of the saddle point $(0)$ and the minimum $(1)$, respectively, and $\Omega_{0}$ is the damped frequency of oscillations about the saddle point $(0)$.  It is given by the absolute value of the unique negative eigenvalue of the transition matrix obtained from the linearized Landau-Lifshitz equation around the saddle point $(0)$.

The application of Eq. (\ref{eq:LangerFormula}) to a single magnetic moment and a standard steepest descents estimate of $Z_{0}$ and $Z_{1}$ leads to Brown's relaxation rate formula \cite{bro79ieee} 
\begin{equation}\label{eq:BrownFormula}
\Gamma_{(1)\rightarrow(2)}\sim\frac{\Omega_{0}\omega_{1}}{2\pi\omega_{0}}e^{-\beta\left(E_{0}-E_{1}\right)},
\end{equation}
 where $\omega_{1}/\omega_{0}=\sqrt{c_{1}^{(1)}c_{2}^{(1)}/(-c_{1}^{(0)}c_{2}^{(0)})}$ is the ratio of frequencies of harmonic oscillations at the minimum $(1)$ and saddle point $(0)$, 
and 
\[
\Omega_{0}=\frac{\beta}{4\tau_{\mathrm{N}}}\left[-c_{1}^{\left(0\right)}-c_{2}^{\left(0\right)}+\sqrt{\left(c_{2}^{\left(0\right)}-c_{1}^{\left(0\right)}\right)^{2}-\frac{4}{\alpha^{2}}c_{1}^{\left(0\right)}c_{2}^{\left(0\right)}}\right].
\]
Here $\alpha$ is the dimensionless damping constant, $\beta=1/k_{B}T$, $k_{B}$ is Boltzmann's constant, $T$ the absolute temperature, $\tau_{\mathrm{N}}=\beta\mu_{s}(1+\alpha^{2})/(2\alpha\gamma)$ is the free diffusion time. 
The coefficients $c_{i}^{\left(j\right)}$ are the Taylor expansion coefficients of the energy (\ref{eq:EnEOSP}) near a stationary point (\textit{j}). In general, the energy possesses several minima and saddle points. The calculation of the \textit{overall} relaxation rate (the inverse relaxation time of the magnetization) must therefore involve a suitable combination of the elementary rates provided by Eq. (\ref{eq:LangerFormula}) or (\ref{eq:BrownFormula}). The escape rate $\Gamma_{(1)\rightarrow(2)}$ in Eqs. (\ref{eq:LangerFormula})-(\ref{eq:BrownFormula}) is valid only in the intermediate to high damping limit, i.e., for \textit{$\alpha$}
$>$ 1. For \textit{$\alpha$}$<$ 1 the relaxation rate should be estimated using a universal turnover formula \cite{Coffey2001}
\begin{equation}
\label{eq:Melnikov}
\Gamma^{'} =A(\alpha S)\Gamma _{(1)\rightarrow (2)} ,
\end{equation}
\bigskip where $A(\alpha S)$\ is the depopulation factor defined as
\begin{equation}
A(\alpha S)=\exp \left[ \frac{1}{\pi }\int_{0}^{\infty }\frac{\ln \{1-\exp\left[ -\alpha S\left( \lambda ^{2}+1/4\right) \right] \}}{\lambda ^{2}+1/4}\,d\lambda\right] 
\end{equation}
and $S$ is the dimensionless action variable. This formula is universal in the sense that it provides an accurate description of the relaxation rate for all values of the damping parameter $\alpha$ \cite{Coffey2001}. 
%
\begin{figure}[ht!]
\begin{center}
\includegraphics[width=4cm,height=4cm]{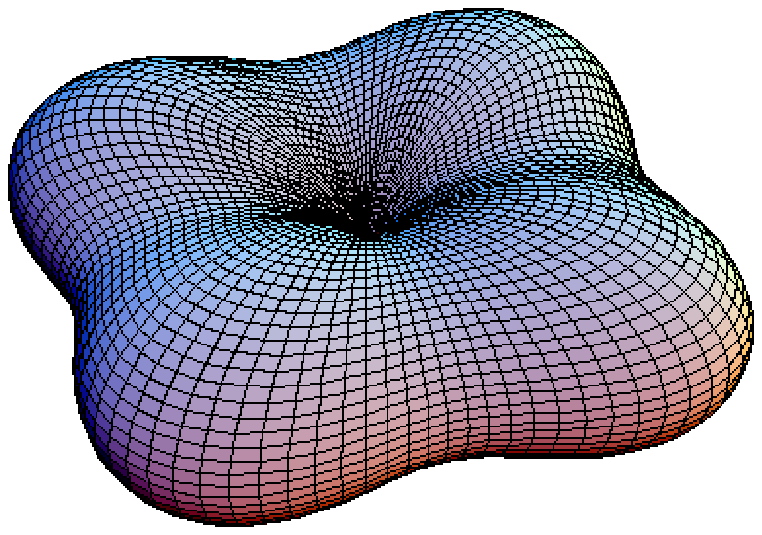}
\includegraphics[width=4cm,height=4cm]{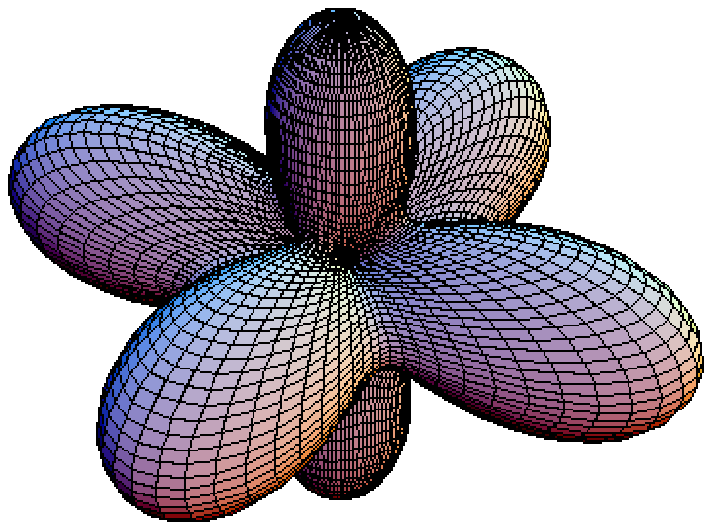}
\end{center}
\caption{\label{fig:EnAllZeta}Energyscape for $\zeta = 1/2$ (left) and $5$ (right) and zero field.}
\end{figure}
%

In the absence of applied field, it is easy to analyze the energyscape  [see Fig. \ref{fig:EnAllZeta}] and obtain analytical expressions for the relaxation rate using Eq. (\ref{eq:LangerFormula}) or (\ref{eq:BrownFormula}). In the case of dominating UA, i.e., for $\zeta<1$, there two minima at $\theta^{(m)}=0,\pi$, with the energy $\varepsilon_{0}^{(m)}=-\sigma(2-\zeta)/2$, and $4$ saddle points at $\theta^{(s)}=\pi/2; \quad\varphi^{(s)} = \pi/4,3\pi/4,5\pi/4,7\pi/4$ whose energy is $\varepsilon_{0}^{(s)}=\sigma\zeta/4$, leading to the energy barrier $\Delta\varepsilon_{0} = \varepsilon_{0}^{(s)} - \varepsilon_{0}^{(m)} = \sigma(4-\zeta)/4$, where $\sigma=\beta K_2V$. As the CA contribution increases there first appear saddle points at the equator (see Fig. \ref{fig:EnAllZeta}), then the minima become flat for $\zeta=1$, and when $\zeta$ is further increased the UA minima split and become local maxima surrounded by $4$ deep minima. For large $\zeta>0$ these minima are located along the directions $\left[\pm 1,\pm 1,\pm 1\right] $.
For $\zeta>1$, the saddles between the regions with $m_{z}>0$ and $m_{z}<0$ and the corresponding energies are the same as for $\zeta<1$, while there are $4$ minima for $m_{z}>0$ and $4$ minima for $m_{z}<0$ that are given by $\cos\theta^{(m)} = \pm\sqrt{(\zeta+2)/3\zeta}, \varphi^{(m)}=\pi/4,3\pi/4,5\pi/4,7\pi/4$, with the corresponding energy $\varepsilon_{0}^{(m)}=\sigma(\zeta^{2}-2\zeta-2)/6\zeta$ leading to the energy barrier $\Delta\varepsilon_{0}=\sigma(\zeta+2)^{2}/12\zeta$. We note that there are also saddle points connecting the minima within each of the regions $m_{z}>0$ and $m_{z}<0$ but they are irrelevant in the calculation of the magnetization reversal time.

Hence, in the absence of applied field, we compute the relaxation rate for each escape route and multiply the result by the symmetry factor ($=8$) that takes into account the number of escape routes. This leads to the following result for the inverse of the reversal time $\Gamma=\tau^{-1}$
\begin{equation}\label{eq:RRAllZeta}
\Gamma\tau_{\mathrm{N}} =\frac{\sigma\mathcal{A}}{\pi\sqrt{2}}\left\{ \begin{array}{ccc}
\frac{\zeta-1}{3\zeta}\, e^{-\sigma(\zeta+2)^{2}/(12\zeta)} & \mathrm{for} & \zeta>1\\ \\
\frac{2(1-\zeta)}{\sqrt{\zeta\left(\zeta+2\right)}}\, e^{-\sigma\left(1-\frac{\zeta}{4}\right)} & \mathrm{for} & \zeta<1,\end{array}\right.
\end{equation}
where
$$
\mathcal{A} \equiv\sqrt{\left(3\zeta+2\right)^{2}+\frac{8}{\alpha^{2}}\zeta\left(\zeta+2\right)}+\left(2-\zeta\right).
$$

For $\zeta\sim1$, the parabolic approximation at the minimum breaks down, so that Eq. (\ref{eq:LangerFormula}) must be used instead of Eq. (\ref{eq:BrownFormula}) as $\textit{Z}_{1}$ can be computed beyond the parabolic approximation. 
Furthermore, it is clear that Eq. (\ref{eq:RRAllZeta}) cannot be used when $\zeta \rightarrow 0$ and an appropriate expression for this crossover to the pure UA has to be used \cite{kachkachietal07prep}.

%
\begin{figure}[ht!]
\begin{center}
\includegraphics[width=8cm]{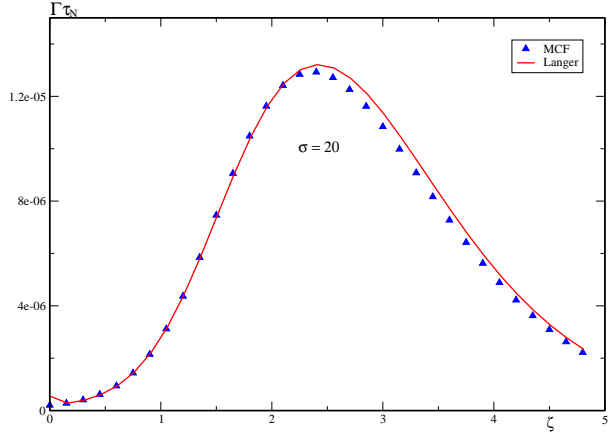}
\end{center}
\caption{\label{fig:eosp_LanVsMCF}Relaxation rate against $\zeta$ for $\sigma=20$ in zero field.}
\end{figure}

The behavior of the relaxation rate in zero field as a function of $\zeta$ is plotted in Fig. \ref{fig:eosp_LanVsMCF} for $\sigma=20$ and $\alpha=1$. The full line is a plot of the formula in Eq. (\ref{eq:LangerFormula}) with $Z_1$ being computed numerically in the whole upper plane of the energyscape [see Fig. \ref{fig:EnAllZeta}]. The triangles represent the result of the MCF calculation. 

When a magnetic field is applied in an arbitrary direction, as is necessary for the calculation of the switching field,  the loci of the stationary points must be found numerically. Accordingly, the relaxation rate can also be calculated numerically using the matrix-continued fraction (MCF) method as suggested by Risken \cite{risken96springer} and developed in Ref. \cite{cofkalwal05worldsc}. According to this method, the solution of the Fokker-Planck equation for the distribution function $W(\vartheta,\phi,t)$ of magnetization orientations may be turned into an eigenvalue problem for which the smallest non vanishing eigenvalue $\lambda_{1}$ yields the relaxation time $\tau$ as $\tau=\lambda_{1}^{-1}$. This relaxation time represents the time of reversal of the magnetization when the ratio $\zeta$ is small, e.g., $\zeta<0.3$. The MCF method provides a systematic check of the asymptotic formulae of the overall escape rate and allows us to bypass the energyscape analysis, which can become rather involved. On the other hand, the analytic calculation can be useful when the barrier is so high that the computation of the relaxation rate by the MCF method becomes extremely time consuming.

As can be seen, good agreement is obtained for all values of $\zeta$. One may see that the relaxation rate reaches a maximum that corresponds to a $\zeta$ value $\zeta_{\max}$. This is due to the fact that for $\zeta<\zeta_{\max}$ the barrier is lowered owing to the creation of saddle points, whereas for $\zeta>\zeta_{\max}$ the barrier is raised owing to a deepening of the minima. As a consequence the escape rate shows a bell-like shape. 
\section{Surface and temperature effects on the switching field}
\begin{figure}[ht!]
\begin{center}
 \includegraphics*[width=7cm, height=7cm]{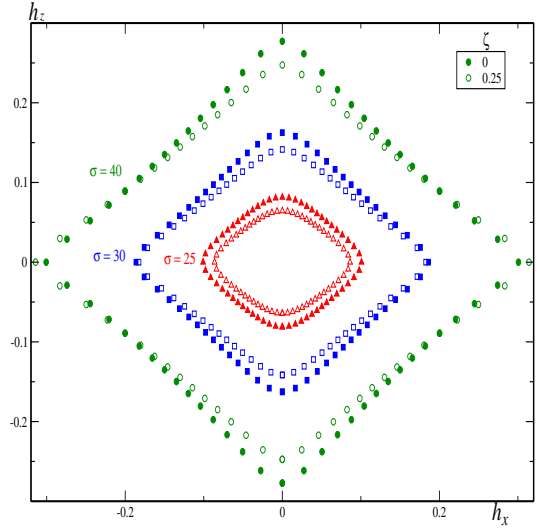}
\end{center}
\caption{\label{fig:2DAstroid}2D switching field for different values of $\sigma$ ($\sim T^{-1}$) and $\zeta$ (cubic anisotropy contribution). The magnetic field is applied in the $xz$ plane and the uniaxial anisotropy axis lies in the $z$ direction.}
\end{figure}

By definition, the switching field (SF) is the field $H_{S}$ at which the magnetization overcomes the energy barrier and reverses its direction. At zero temperature, this reversal is only possible if the energy barrier is fully suppressed.
This requires the application of an external magnetic field, called the critical field or the field at the limit of metastability. This definition corresponds to the SW critical field whose extremity describes the SW astroid, i.e., the limit-of-metastability curve \cite{wernsdorfer01acp}.
At finite temperatures, the SF is intuitively smaller and relaxation of the magnetization must be accounted for. Moreover, \textit{experimental} observation of the magnetization reversal depends on the relaxation time of the cluster \textit{and} on the measuring time $\tau_{m}$ of the experimental setup. Therefore the magnetization reversal can be \textit{experimentally} observed only if the
relaxation time is \textit{in the time window of the experiment}, or equivalently, if the relaxation rate is equal to the measuring frequency $\nu_{m}=\tau_{m}^{-1}$. Therefore, for the experimental observation of the magnetization reversal at finite temperatures, we must have 
\begin{equation}\label{eq:SwitchingCondition}
\Gamma\left(\sigma,\zeta,h_{S},\psi,\alpha\right)=\nu_{m}
\end{equation}
where $h_{S}=\mu_{s}H_{S}/(2K_2V)$ is the switching field to anisotropy field ratio. This equation can be solved numerically for $h_{S}\left(\sigma,\zeta,\psi,\alpha\right)$ and $\nu_{m}$ for given values of $\sigma$, $\zeta$, $\alpha$ and $\psi$. At very low temperatures ($\sigma\rightarrow\infty$) and $\zeta=0$ this yields the SW astroid \cite{wernsdorfer01acp} . For the given parameters $\sigma,\zeta,\alpha,\nu_{m}$, $h_{S}$ can be plotted against the angle $\psi$. For small $\zeta$ values,  a parametric plot of the components $h_{x}=h_{S}\sin\psi,h_{z}=h_{S}\cos\psi$, with $h_{S}$ obtained from Eq. (\ref{eq:SwitchingCondition}) may be used. The field-dependent reversal time $\tau$ is computed using the MCF technique for different values of the parameters $\sigma\sim T^{-1}$ and $\zeta$  and the result is plotted in Fig. \ref{fig:2DAstroid}.
%
%
 In fact, we calculate the dimensionless relaxation rate $\Gamma\tau_{\mathrm{N}}$ with $\tau_{\mathrm{N}}=\tau_{s}\times\sigma$, where $\tau_{s}=1/\left(\alpha\gamma H_{2}\right)$ with $H_{2}=2K_2V/\mu_{s}$, determines the time scale within the system.
For a spherical cobalt cluster of $3\,\mathrm{nm}$ in diameter with $\alpha=1.0$, $\tau_{s}$ evaluates to $\tau_{s}\simeq1.7636\times10^{-11}\,\mbox{s}$. In Fig. \ref{fig:2DAstroid} the measuring frequency has been taken as $\nu_{m}=100\,\mathrm{Hz}$.

Two observations can be made from the results in Fig. \ref{fig:2DAstroid}. First, we observe a nearly homogeneous shrinking of the SF curve due to thermal fluctuations, and this has been observed in cobalt nanoclusters in Ref. \cite{jametetal01prl} and also obtained numerically in Ref. \cite{vouilleetal04jmmm} within the Langevin approach. 
The SF decreases with temperature owing to a decrease in the relaxation time. Moreover, Fig. \ref{fig:2DAstroid} shows that upon increasing the CA contribution, i.e., when $\zeta$ goes from $0$ to $0.25$, there is a flattening of the SF curve, that is the critical field is reduced in the longitudinal direction ($\psi=0$) and increased in the transversal direction ($\psi=\pi/2$). Recalling that $\zeta$ (or $K_{4}$) is related to the SA, these results show how the latter affects the switching field. 
This flattening of the SF curve owing to SA was observed in Ref. \cite{kacdim02prb} where such a curve was computed at very low temperature for a spherical many-spin particle with local (on-site) anisotropy, uniaxial in the core and transverse on the surface with a variable constant.

In the present work, we find that when the SA changes in intensity, the relaxation rate changes by more than an order of magnitude. More precisely, Fig. \ref{fig:EnAllZeta} shows that when $\zeta = K_4/K_2$ increases from (nearly) zero (dominating UA) and exceeds $2$, the (dimensionless) relaxation rate $\Gamma\tau_{\mathrm{N}}$ increases from its minimum ($\sim 10^{-6}$) and exceeds $10^{-5}$. 
Moreover, we note that since the relaxation rate has a maximum, optimum stability of the cluster magnetization against thermally-activated reversal is better achieved either at very small or large $\zeta$. Indeed, for very small $\zeta$ (dominating UA) there are no saddle points, and for large $\zeta$ (dominating CA) the energy barriers are very high, and thereby the relaxation rate, or the switching probability, is very small.
In this context, these results imply that one may tune the thermal stability of the nanocluster magnetization by tuning its SA contribution.

These effects can be observed in $\mu$-SQUID measurements on isolated (Co or CoPt) clusters especially at low temperature, and as a matter of fact, some preliminary measurements have indeed confirmed the flattening of the switching field curve but more measurements are needed to confirm this result.
\section{Conclusions}
The magnetization reversal time of a magnetic nanocluster is calculated taking account of surface anisotropy by using an effective macrospin model \cite{EOSP} that maps the energyscape of the many-spin nanocluster onto that of a macrospin with uniaxial and cubic anisotropies. The mapping allows us to compute the relaxation rate of a magnetic moment in the effective anisotropy potential as a function of the cubic-to-uniaxial anisotropy ratio so that surface anisotropy is accounted for. The analytical asymptotic formulae for the reversal time are given in both limits of weak and strong cubic anisotropy in the absence of applied field. The cubic anisotropy first creates saddle points, thus lowering the energy barrier, and then by further embedding of the minima it raises the energy barrier. The matrix-continued fractions method is used in order to check the range of validity of the asymptotic expressions of the relaxation rate in zero field.  This method is also used to compute the reversal time for an arbitrary magnitude and direction of the field in order to calculate the switching field curves. 

These developments allow us to compute the switching field curves for weak cubic anisotropy as a function of temperature and intensity of the cubic anisotropy contribution. The switching field curve exhibits two main features. First, this curve homogeneously shrinks with increasing temperature and disappears completely at the blocking temperature of the nanocluster. This is in a qualitative agreement with the experimental observations on cobalt nanoclusters.
Second, at a given temperature, we observe a flattening of the switching field curve when the intensity of the cubic anisotropy increases. More precisely, the longitudinal component of the switching field decreases whereas its transverse component increases. The more involved case of strong cubic anisotropy is left for a future investigation. Finally, we emphasize that a comparison between the present results and $\mu$-SQUID measurements would provide us with an approach of further investigating surface anisotropy in magnetic nanoclusters.

We thank William T. Coffey for reading the manuscript and making useful comments. HK acknowledges instructive discussions with V\'eronique Dupuis about the $\mu$-SQUID measurements.

\section*{References}

\end{document}